# A Framework for Uncertainty Estimation in Seismology Data Processing with Application to Extract Rayleigh Wave Dispersion Curves from Noise Cross-correlation Functions

Ziye Yu[1], Xin Liu[2,3]

1 Institute of Geophysics, China Earthquake Administration, China;
2 Laboratory of Seismology and Physics of Earth's Interior, School of Earth and Space Sciences, University of Science and Technology of China, Hefei 230026, China;
3 Institute of Advanced Technology, University of Science and Technology of China, Hefei 230088, China.

Abstract: Extracting meaningful information from large seismic datasets often requires estimating the uncertainty associated with the results for quantitative analysis. This uncertainty arises from both the raw data and the manually labeled annotations. We introduce an uncertainty estimation framework designed to calculate the uncertainty from manually labeled data. This framework can efficiently output the true posterior from large datasets. We apply the framework to extract Rayleigh wave phase velocity dispersion and compute the posterior distribution of the dispersion results. We utilize 62,899 noise cross-correlation function (NCF) data from 438 stations located in Yunnan Province and manually label the Rayleigh phase velocity dispersion curves. Dispersion curve extraction presents two key challenges: (1) Researchers typically derive dispersion curves from spectrograms in the period–velocity domain, limiting the ability to directly study the relationship between NCFs and dispersion curves; (2) Assessing uncertainty in manually labeled data remains difficult. To address these challenges, the framework takes the NCFs as input and directly output both the dispersion values and the posterior of the dispersion values when processing the NCF data. This approach allows us to construct a flexible deep neural network (DNN) architecture that balances accuracy and computational efficiency.

## 1 Introduction

There are many techniques for extracting information from seismic records. For example, we can determine the Pg/Sg arrival times from three-component waveform data(Mousavi et al., 2020; Yu et al., 2023; Yu & Wang, 2022; Zhu & Beroza, 2018), or we can extract phase/group velocity dispersion curves from dispersion spectrograms (Dong et al., 2021; Song et al., 2022; S. Yang et al.,

2022). These tasks require manually labeling data to train deep neural networks (DNN) or any other machine learning models. However, errors can occur when manually labeling data, and it is challenging to access these errors or uncertainties. In this paper, we demonstrate how to perform efficient approximate posterior inference with arbitrary DNN architectures and optimize the models using general stochastic gradient ascent techniques. We use the task of extracting Rayleigh wave dispersion curves as an example.

Over the past decade, the ambient noise tomography has greatly increased our understanding of crust and upper mantle(Bensen et al., 2009; Fang et al., 2015; Li et al., 2022; Liu et al., 2023; Nimiya et al., 2023). In early studies, surface waves were widely used to investigate velocity structures on a global scale(Bensen et al., 2008; Y. Yang et al., 2007). Due to the layout of seismic arrays and the need to observe local shallow crustal structures, short-period surface waves ranging from 3 to 50 seconds have been extensively utilized (Liu et al., 2023). The inversion process using ambient noise cross-correlation surface waves generally involves two main steps (Fang et al., 2015): first, extracting dispersion curves from the ambient noise cross-correlation functions; second, inverse the 3D velocity from the dispersion curves. Extracting dispersion curves has become the key step in ambient noise tomography.

Fundamental-mode Rayleigh and Love phase/group velocity dispersion curves are widely utilized in seismic tomography (S. Yang et al., 2022; Yao et al., 2005, 2006, 2011). The extraction of dispersion curves is typically based on either single-station or two-station analysis (Dong et al., 2021), while ambient noise cross-correlation requires computing surface wave signals between arbitrary pairs of stations. Traditionally, dispersion curve extraction from noise cross-correlation functions (NCFs) relies on Frequency-Time Analysis (FTAN) methods to convert the NCF into a period–velocity spectrogram image (Yao et al., 2005, 2006, 2011). Once a high-quality spectrogram is obtained, the peak energy values are extracted to derive the dispersion curve. However, manually picking dispersion curves from such images is a labor-intensive process.

To address this issue, recent studies have leveraged deep neural networks (DNNs) to automate dispersion curve extraction from spectrogram images. Zhang et al. (2020) employed U-Net model to extract group velocity curves from FTAN images. Dong et al. (2021) utilized U-Net-like model to extract multimode dispersion curves from Frequency-Bessel dispersion spectrum images. Similarly, S. Yang et al. (2022) applied a U-Net to extract fundamental mode of group and phase velocity dispersion from phase and group spectrograms. Most of these methods use FTAN images as input but do not provide uncertainty estimates when extracting dispersion curves.

Thus, the task of extracting dispersion curves serves as an excellent example to validate our uncertainty estimation framework. Our framework can utilize arbitrary DNN architectures to meet the requirements of uncertainty estimation. A large-scale network can achieve higher accuracy and provide physical insights for dispersion curve extraction, while a medium-scale network offers a balance between accuracy and inference speed. The application to the ChinArray X1 dataset demonstrates that our framework can generate reliable dispersion curves along with a true posterior distribution of the dispersion curve.

## 2 Method

### 2.1 The uncertainty estimation framework

We want to estimate the uncertainty of the dispersion $d$ when processing NCF data $x$:

$$p(d|x) \tag{1}$$

However, the posterior distribution of dispersion $d$ is intractable. While Markov Chian Monte Carlo (MCMC) methods methods (Li, 2021) can be used to estimate the posterior distribution, they involve computationally expensive iterative inference schemes (Kingma & Welling, 2022). To address this, we aim to optimize the models using general stochastic gradient ascent techniques. Therefore, we introduce a latent variable $z$ to generate the dispersion $d$, and maximum the evidence lower bound (ELBO):

$$\log p(d|x) \geq \int_z q(z|x) \log\left(\frac{p(z,d|x)}{q(z|x)}\right) dz = ELBO \tag{2}$$

$$ELBO = \int_z q \log(p(z|x)p(d|z,x)) dz - \int_z q \log q = \int_z q \log \frac{p}{q} dz + \int_z q \log p(d|z,x) dz \tag{3}$$

Let the prior of latent variables be the standard normal distribution $p(z|x) = \mathcal{N}(z; 0,1)$. The variational distribution $q(z|x) = \mathcal{N}(z, \mu, \sigma)$ is parameterized as a normal distribution. Then the first term can be described as:

$$\int_z q \log \frac{p}{q} dz = \frac{1}{2}\sum_{j=1}^{J}(1 + \log(\sigma_j^2) - \mu_j^2 - \sigma_j^2) \tag{4}$$

The second term of ELBO can be:

$$\int_z q \log p(d|z,x) dz \simeq \frac{1}{N}\sum_{i=1}^{N} \log p(d|z^s) \tag{5}$$

Where $z^s$ is sampled from $q$, $J$ is the length of vector $z$. We can utils a DNN to model $p(d|z,x) \rightarrow d = DNN(z,x)$.

There are two types of input: the raw NCF data $x$ with 1 Hz sampling rate; the noise data. There are two types of output $d$: the classification output $y^c \in \mathbb{R}^K$ and regression output $y^r \in \mathbb{R}^K$, where $K$ is the number of periods that are marked by human. The regression output $y_k^r$ indicate the velocity value on the k-th period. The classification output $y_k^c$ is used to mark the dispersion $y_k^r$ on the k-th period is available, which 1 is available and 0 is filtered out. Then the overall structure of the input and output of the DNN model can be described as:

$$y^r, y^c = DNN(x,z) \tag{6}$$

The DNN model can be arbitrary structure, whose inputs are $x, z$ and outputs are $y^r, y^c$. The $ELOB$ can be:

$$ELBO = \frac{1}{N}\sum_{i=1}^{N} \log p(d|z^s) + \sum_{j=1}^{J}(1 + \log(\sigma_j^2) - \mu_j^2 - \sigma_j^2)$$

$$= mean((y^r - d^r)^2 \cdot y^d + d^c \log y^c + (1 - d^c) \log(1 - y^c)) + \sum_{j=1}^{J}(1 + \log(\sigma_j^2) - \mu_j^2 - \sigma_j^2)$$

(7)

Where $d^r, d^c$ is the target label of regression and classification.

We build multi-layer Transformer as the DNN model. Shown in Figure 1 :

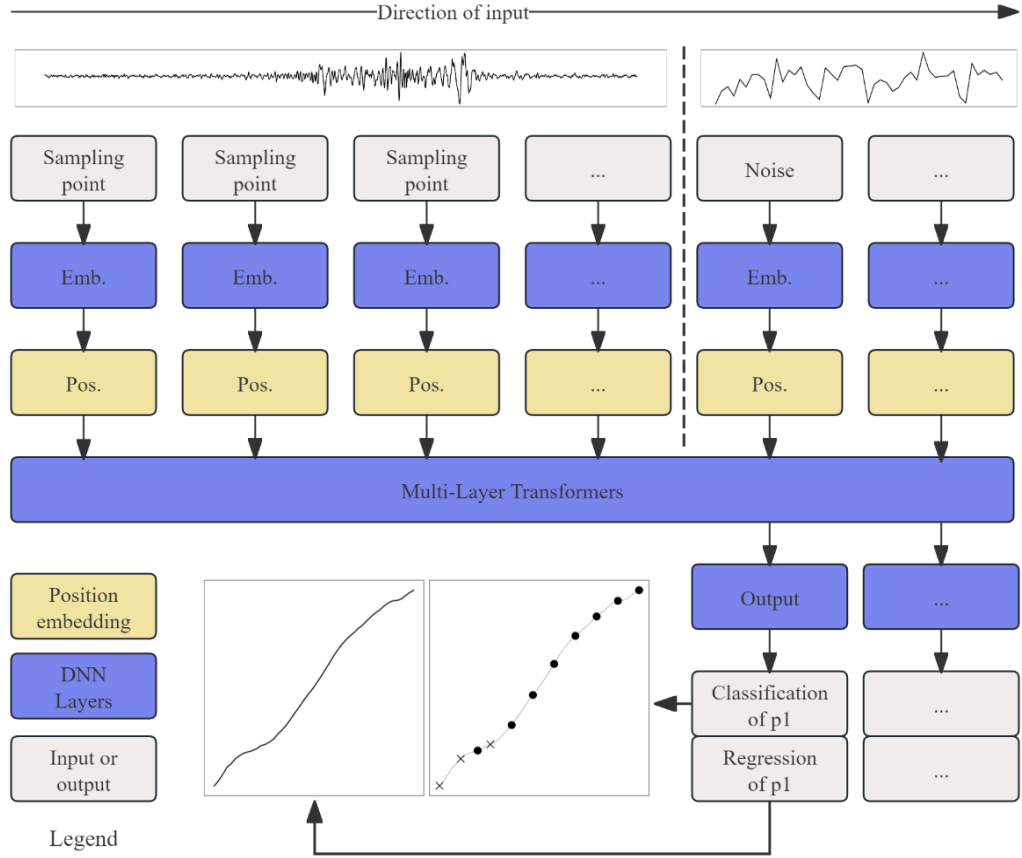

Figure 1  The architecture of DNN models.

2.2 Dataset

In this study, we used data from the first phase of the ChinArray project and permanent seismic stations, totaling 438 stations, deployed by the Institute of Geophysics, China Earthquake Administration, between 2011 and 2013 in the northern segment of the North-South Seismic Belt. These stations provided dense coverage across the study region. The ambient noise cross-correlation functions (NCFs) were calculated using the method proposed by Benson et al., and dispersion curves were manually labeled following the method of EGFAnalysisTeimFreq (Yao et al., 2005, 2006, 2011). A total of 62,899 usable NCFs with corresponding labeled dispersion data were obtained. The station distribution is shown in Figure 2 (a), illustrating a relatively uniform coverage; all NCFs are displayed in Figure 1(b), with most inter-station distances being less than 1000 km.

For the cross-correlation function between two stations, A and B, the positive time axis can be regarded as surface waves propagating from A to B, while the negative time axis corresponds to propagation from B to A. In conventional processing, the positive and negative time axes are stacked together. However, since the waveform characteristics of the positive and negative halves are not entirely identical, we adopted two input data processing strategies: (1) The traditional method of stacking the positive and negative time axes, referred to in this paper as the **stacked dataset**; (2) Separately using the positive and negative halves along with their respective labels as independent training samples, thereby doubling the number of training data compared to the stacked dataset. This is referred to as the single-side dataset. All input NCFs were uniformly truncated to a length of 1024 samples, corresponding to 1024 seconds. To minimize preprocessing overhead, we applied only minimal processing to the NCFs, which included detrending and normalization (standardization). Since simulated or augmented data may not closely represent real data, and the manually labeled dataset is sufficiently large, we did not perform any data augmentation or introduce synthetic data.

The manually labeled dispersion points from the NCFs were organized into two types of labels (Figure 3 ). The classification output to predict which frequency points are usable. The regression output to predict the velocity values for the usable frequency points. The dispersion labels were annotated over the period range of 3–50 seconds, with one frequency point labeled at each 1-second interval.

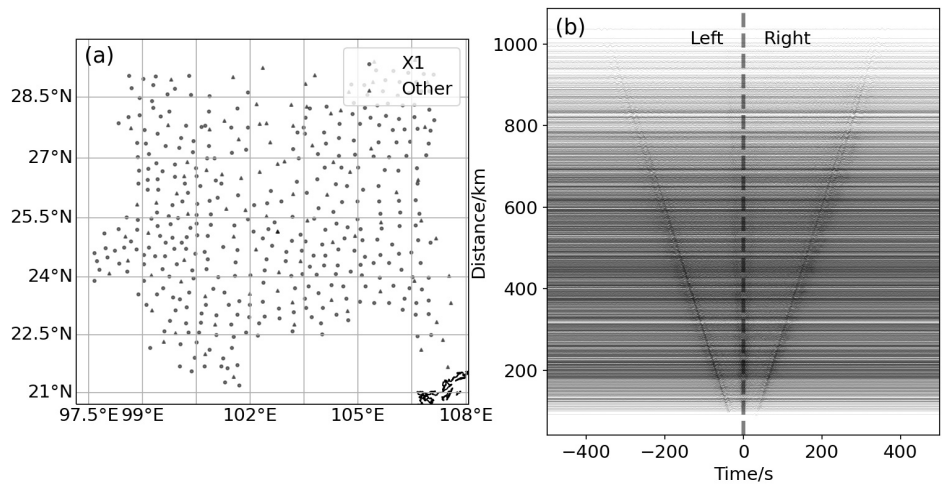

Figure 2 Distribution of seismic stations and cross-correlation data used for model training and testing.
(a) Includes permanent stations (triangles) and mobile stations (dots), totaling 438 stations.
(b) Shows 62,899 manually labeled NCFs.

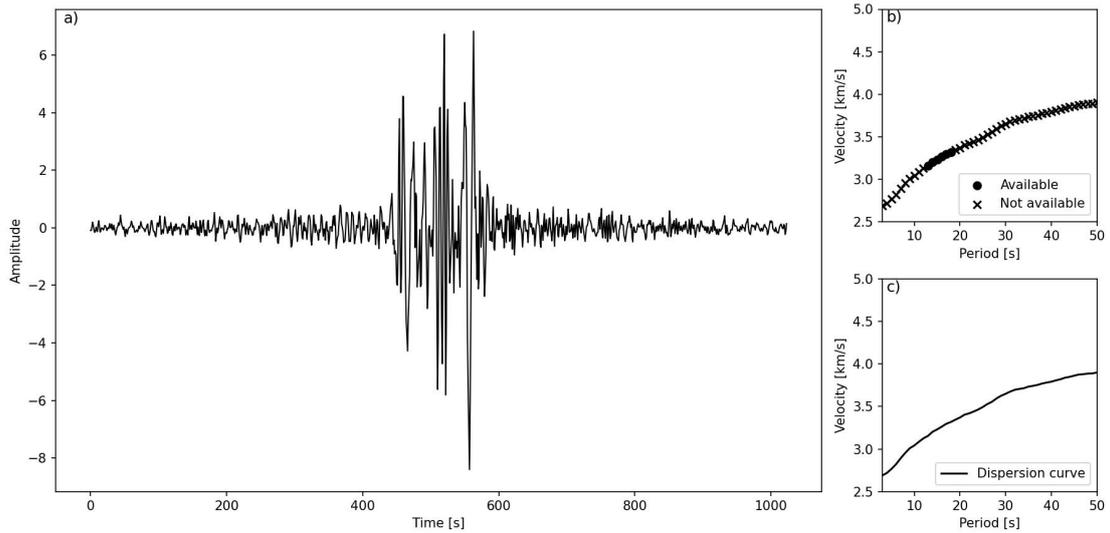

Figure 3 The NCF and the label of classification and regression label
**(a)** NCF waveform (input to the model);
**(b)** Output from the classification model (indicating usable frequency points);
**(c)** Output from the regression model (predicted velocity values at each frequency point).

For model training, the dataset was divided into a training set and a test set. The training set was used to train the model, while the test set was used to evaluate the model's generalization capability. To ensure independence between the two sets and to make the test set representative of real-world application data, we selected NCFs in which at least one of the two stations was a permanent station as the training set, totaling 24,360 samples. The remaining data were used as the test set. By using a relatively small training dataset, we aimed to better reflect the model's performance on real data.

## 3 Results and Testing

### 3.1 The prediction result

The model can directly output the classification and regression result Figure 4 .

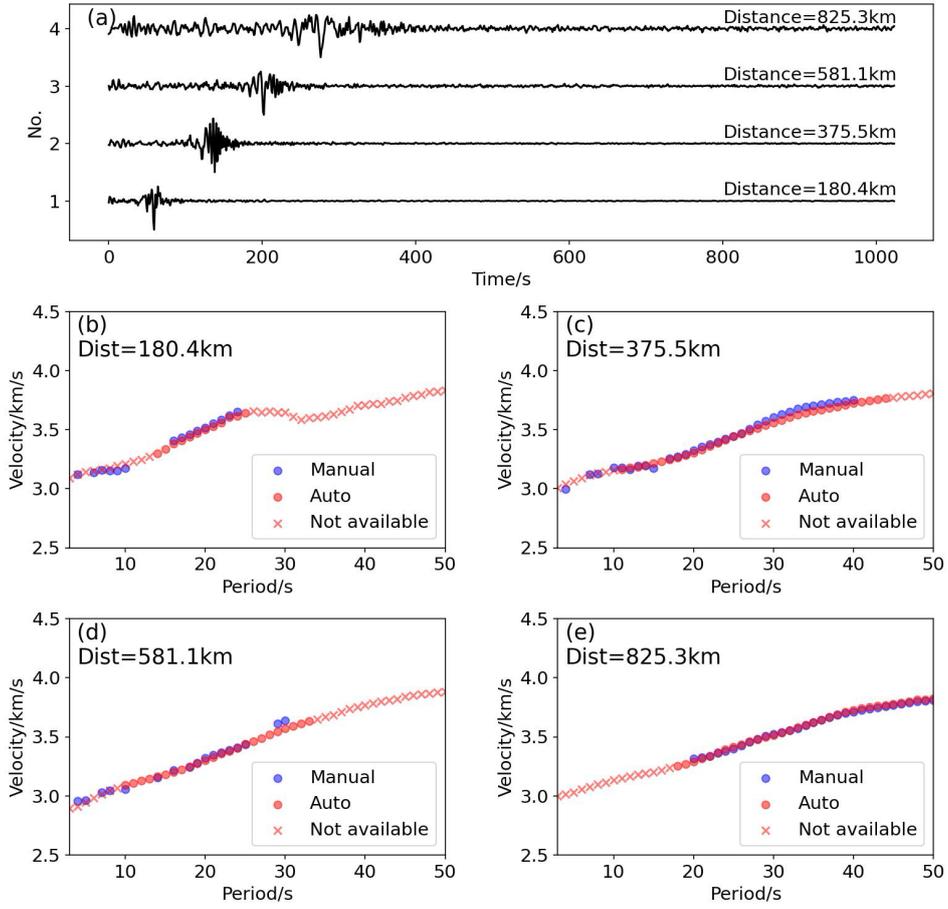

Figure 4  Comparison Between Manual Labels and Automatic Picking Results
(Four Random Samples)
(a) NCF curves used for testing;
(b)–(e) Automatic picking results for four samples.

From the results, we observe that the classification outcomes (i.e., available or not available, as shown in Figure 4 ) closely resemble the manually labeled data. This indicates that the DNN model successfully learns the manual quality control scheme. Additionally, the available data exhibit greater continuity compared to the labeled data. The regression results closely match the manually labeled dispersion curve values, demonstrating that our model can directly output dispersion curves without relying on FTAN data. To further estimate the posterior distribution, we sample random data $z \sim \mathcal{N}(\mu^*, \sigma^*)$, where the $\mu^*, \sigma^*$ are derived from the training process, and then compute the statistical distribution of the dispersion results. The resulting distribution is shown in Figure 5

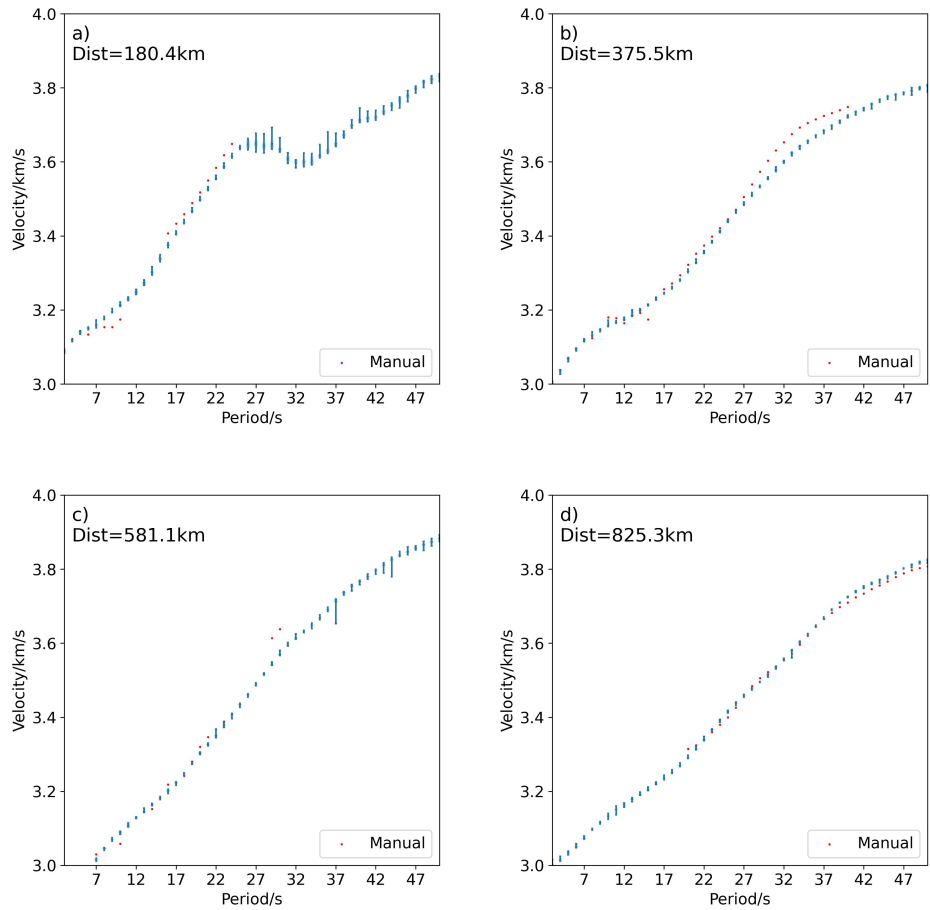

Figure 5  The violin plot of the distribution predicted by DNN model.

The distribution output by the DNN represents the true posterior, whereas most previous studies approximate the distribution as a Gaussian distribution. When the station distance is small, the predicted dispersion values at longer periods exhibit relatively larger uncertainty, which is consistent with the underlying physical mechanisms.

## 3.2 Quantitative analysis

To evaluate model performance, we compared the accuracy of models trained on single-side data. The testing was conducted on three types of data: (1) Stacked positive and negative NCF halves; (2) The side with stronger signal amplitude; (3) The side with weaker signal amplitude.

Table 1  The statics on classification and regression result.

| Test dataset | Precison | Recall | Mean error [km/s] | Standard deviation [km/s] |
|---|---|---|---|---|
| Stacked data | 0.736 | 0.805 | 0.0056 | 0.0501 |
| Stronger data | 0.755 | 0.769 | 0.0117 | 0.0510 |
| Weaker Data | 0.747 | 0.735 | 0.0105 | 0.0620 |

We can see from Table 1  that the stacked dataset has higher precision. We also plotted the error distribution for the model trained on single-side data and test on three types of datasets, as shown in Figure 6 .

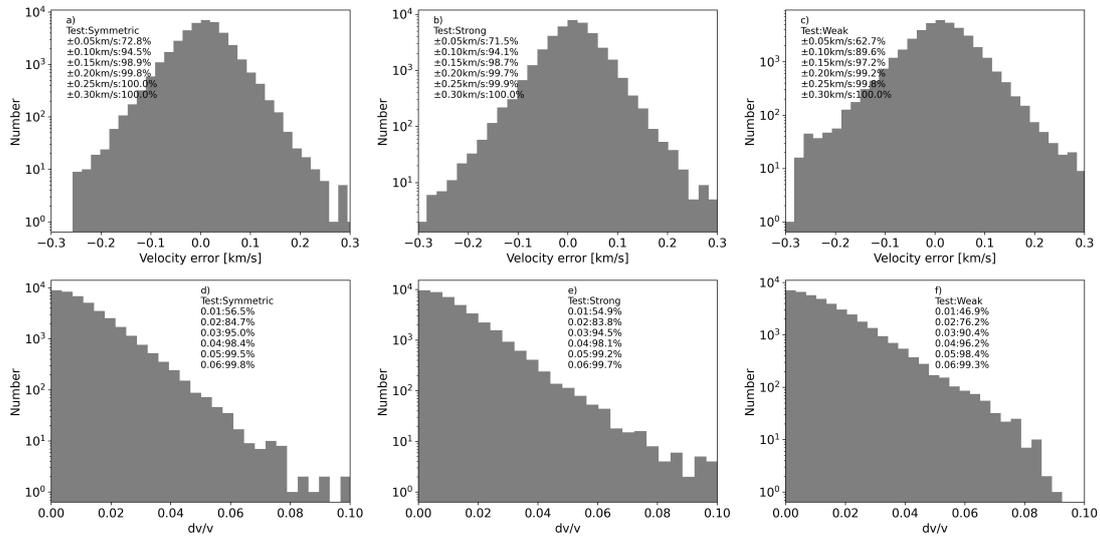

Figure 6  The error distribution of different test data.

We can see from Figure 6 That 95.0% error are smaller than 3% on stacked data. The strong data has similar accuracy with stacked data. The weak data has lower accuracy. That means that we can use the tradition data, which are stacked, to get a moderate accuracy.

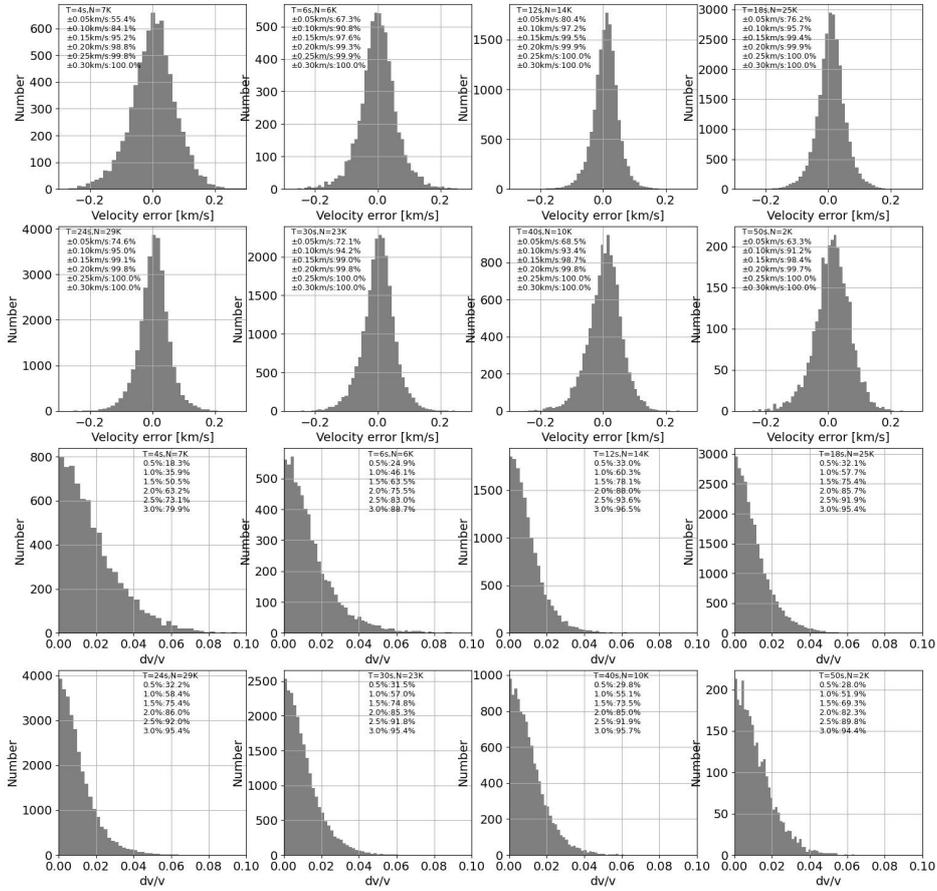

Figure 7  The error distribution on different period.

It can be observed from Figure 7 that the picking accuracy at 4 and 6 seconds is relatively low, whereas it improves significantly at longer periods. This is likely due to the lower signal-to-noise ratio at shorter periods and the higher reliability of dispersion signals at longer periods.

## 4 Discussion

We can also obtain the error distribution by adopting a strategy of adding noise to the data. To verify this approach, we trained a model without a latent variable and introduced Gaussian random noise with a standard deviation of 0.3 to generate different predictions. The results are shown in Figure 8 .

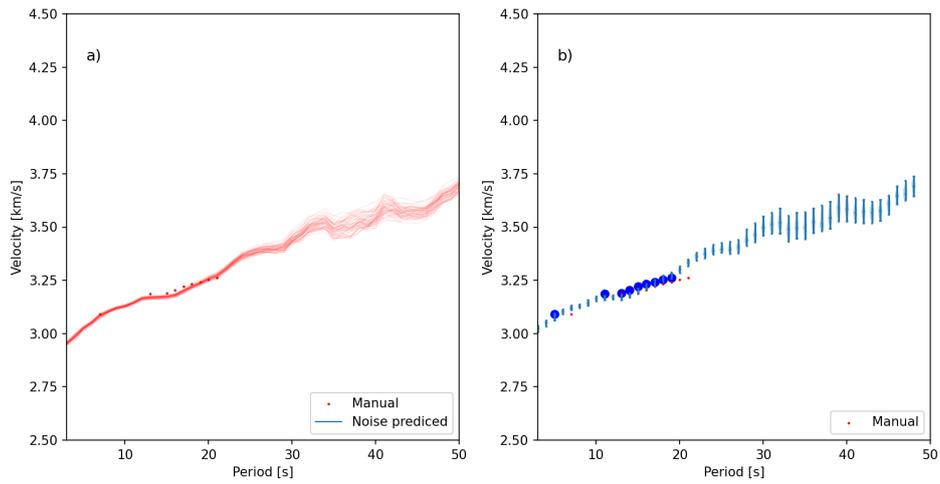

Figure 8  Strategy of adding noise to obtain the distribution

(a) different prediction result;

(b) the violin plot of different periods.

We observe that the distribution across different periods is consistent with the manually labeled data, where periods without manual labels exhibit higher uncertainty. However, since the true noise level of the NCF is unknown, it is challenging to add appropriate noise. Different noise levels added to the NCF data yield varying distribution results. Furthermore, the similarity in uncertainty across different periods suggests that the obtained distribution does not represent the true posterior.

The model is built by Transformer, that means that we can see the relationship between the dispersion curve point and the NCF sampling point (Figure 9).

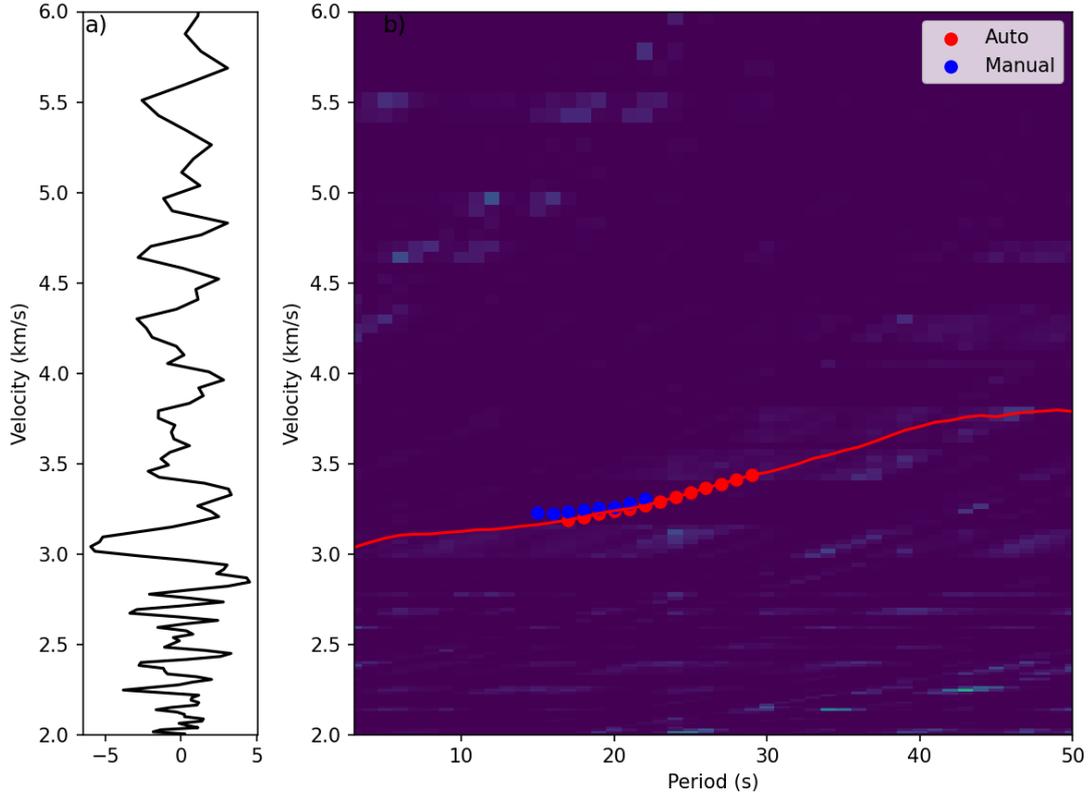

Figure 9  The attention result when extracting dispersion curve.

In Figure 9, we focus on the first attention map, which is closely related to the NCF data. The distribution of attention hotspots follows a trend similar to the dispersion curve. However, some hotspots appear outside the dispersion curve, indicating that the DNN utilizes additional waveform information beyond the surface wave region to extract dispersion characteristics. Since the distance information is not provided as input to the network, the DNN relies on additional waveform characteristics to identify the surface wave segment.

## 5 Conclusion

We developed a novel framework for estimating the posterior distribution in seismic data processing. The framework demonstrates strong performance in extracting surface wave dispersion. The key advantages of our framework are as follows:

1. It allows the construction of arbitrary DNN architectures for data processing.

2. It enables the use of stochastic gradient descent methods to estimate the posterior distribution, which is computationally more efficient than McMC methods.

3. The framework is adaptable to various seismic data processing tasks with minimal modifications to the original network, such phase picking.

4. The dispersion extraction model is end-to-end, allowing us to obtain additional insights, such as attention mechanisms, from the network.

Uncertainty in inversion arises from both physical mechanisms and data uncertainty. Our framework effectively estimates data uncertainty. In future work, we aim to develop new methods to quantify uncertainty related to physical mechanisms. The test code is open source: https://github.com/cangyeone/dispersion

## Reference

Bensen, G. D., Ritzwoller, M. H., & Shapiro, N. M. (2008). Broadband ambient noise surface wave tomography across the United States. *Journal of Geophysical Research: Solid Earth*, *113*(B5), 2007JB005248. https://doi.org/10.1029/2007JB005248

Bensen, G. D., Ritzwoller, M. H., & Yang, Y. (2009). A 3-D shear velocity model of the crust and uppermost mantle beneath the United States from ambient seismic noise. *Geophysical Journal International*, *177*(3), 1177–1196. https://doi.org/10.1111/j.1365-246X.2009.04125.x

Dong, S., Li, Z., Chen, X., & Fu, L. (2021). DisperNet: An Effective Method of Extracting and Classifying the Dispersion Curves in the Frequency–Bessel Dispersion Spectrum. *Bulletin of the Seismological Society of America*, *111*(6), 3420–3431. https://doi.org/10.1785/0120210033

Fang, H., Yao, H., Zhang, H., Huang, Y.-C., & Van Der Hilst, R. D. (2015). Direct inversion of surface wave dispersion for three-dimensional shallow crustal structure based on ray tracing: Methodology and application. *Geophysical Journal International*, *201*(3), 1251–1263. https://doi.org/10.1093/gji/ggv080

Kingma, D. P., & Welling, M. (2022). *Auto-Encoding Variational Bayes* (No. arXiv:1312.6114). arXiv. https://doi.org/10.48550/arXiv.1312.6114

Li, Z. (2021). A Review of Bayesian Posterior Distribution Based on MCMC Methods. In W. Cao, A. Ozcan, H. Xie, & B. Guan (Eds.), *Computing and Data Science* (Vol. 1513, pp. 204–213). Springer Nature Singapore. https://doi.org/10.1007/978-981-16-8885-0_17

Li, Z., Shi, C., Ren, H., & Chen, X. (2022). Multiple Leaking Mode Dispersion Observations and Applications From Ambient Noise Cross-Correlation in Oklahoma. *Geophysical Research Letters*, *49*(1), e2021GL096032. https://doi.org/10.1029/2021GL096032

Liu, Y., Yu, Z., Zhang, Z., Yao, H., Wang, W., Zhang, H., Fang, H., & Fang, L. (2023). The high-resolution community velocity model V2.0 of southwest China, constructed by joint body and surface wave tomography of data recorded at temporary dense arrays. *Science China Earth Sciences*, *66*(10), 2368–2385. https://doi.org/10.1007/s11430-022-1161-7

Mousavi, S. M., Ellsworth, W. L., Zhu, W., Chuang, L. Y., & Beroza, G. C. (2020). Earthquake transformer—An attentive deep-learning model for simultaneous earthquake detection and phase picking. *Nature Communications*, *11*(1), 3952. https://doi.org/10.1038/s41467-020-17591-w

Nimiya, H., Ikeda, T., & Tsuji, T. (2023). Multimodal Rayleigh and Love Wave Joint Inversion for S-Wave Velocity Structures in Kanto Basin, Japan. *Journal of Geophysical Research: Solid Earth*, *128*(1), e2022JB025017. https://doi.org/10.1029/2022JB025017

Song, W., Feng, X., Zhang, G., Gao, L., Yan, B., & Chen, X. (2022). Domain Adaptation in Automatic Picking of Phase Velocity Dispersions Based on Deep Learning. *Journal of Geophysical Research: Solid Earth*, *127*(6), e2021JB023389. https://doi.org/10.1029/2021JB023389

Yang, S., Zhang, H., Gu, N., Gao, J., Xu, J., Jin, J., Li, J., & Yao, H. (2022). Automatically Extracting Surface-Wave Group and Phase Velocity Dispersion Curves from Dispersion Spectrograms Using a Convolutional Neural Network. *Seismological Research Letters*, *93*(3), 1549–1563. https://doi.org/10.1785/0220210280

Yang, Y., Ritzwoller, M. H., Levshin, A. L., & Shapiro, N. M. (2007). Ambient noise Rayleigh wave tomography across Europe. *Geophysical Journal International*, *168*(1), 259–274. https://doi.org/10.1111/j.1365-246X.2006.03203.x

Yao, H., Gouédard, P., Collins, J. A., McGuire, J. J., & Van Der Hilst, R. D. (2011). Structure of young East Pacific Rise lithosphere from ambient noise correlation analysis of fundamental- and higher-mode Scholte-Rayleigh waves. *Comptes Rendus. Géoscience*, *343*(8–9), 571–583. https://doi.org/10.1016/j.crte.2011.04.004

Yao, H., Van Der Hilst, R. D., & De Hoop, M. V. (2006). Surface-wave array tomography in SE Tibet from ambient seismic noise and two-station analysis—I. Phase velocity maps. *Geophysical Journal International*, *166*(2), 732–744. https://doi.org/10.1111/j.1365-246X.2006.03028.x

Yao, H., Xu, G., Zhu, L., & Xiao, X. (2005). Mantle structure from inter-station Rayleigh wave dispersion and its tectonic implication in western China and neighboring regions. *Physics of the Earth and Planetary Interiors*, *148*(1), 39–54. https://doi.org/10.1016/j.pepi.2004.08.006

Yu, Z., & Wang, W. (2022). LPPN: A Lightweight Network for Fast Phase Picking. *Seismological Research Letters*, *93*(5), 2834–2846. https://doi.org/10.1785/0220210309

Yu, Z., Wang, W., & Chen, Y. (2023). Benchmark on the accuracy and efficiency of several neural network based phase pickers using datasets from China Seismic Network. *Earthquake Science*, *36*(2), 113–131. https://doi.org/10.1016/j.eqs.2022.10.001

Zhang, X., Jia, Z., Ross, Z. E., & Clayton, R. W. (2020). Extracting Dispersion Curves From Ambient Noise Correlations Using Deep Learning. *IEEE Transactions on Geoscience and Remote Sensing*, *58*(12), 8932–8939. https://doi.org/10.1109/TGRS.2020.2992043

Zhu, W., & Beroza, G. C. (2018). PhaseNet: A Deep-Neural-Network-Based Seismic Arrival Time Picking Method. *Geophysical Journal International*. https://doi.org/10.1093/gji/ggy423

## Appendix

The CNN architecture used for extracting dispersion curve

We have tested a CNN model to get the posterior of the model. The result can be seen from Figure 10 .

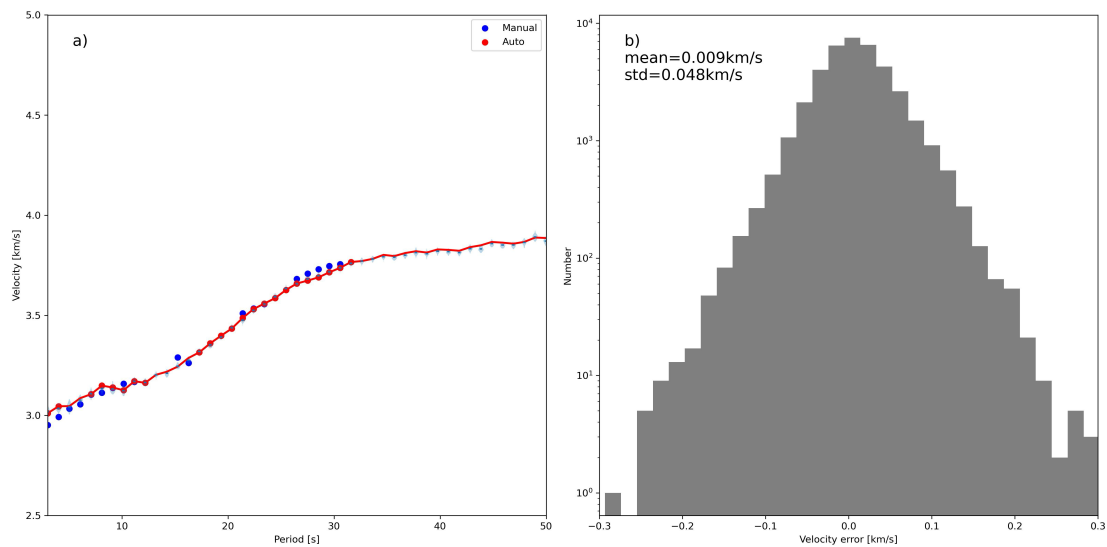

Figure 10  The test accuracy of CNN model.

(a) the prediction results of CNN; (b) the error distribution.

We observe that the CNN model achieves results comparable to those of the Transformer model, with slightly higher accuracy. However, unlike the Transformer model, the CNN model does not provide an interpretable attention mechanism, making it more challenging to explain its decision-making process.